\documentclass[prl,twocolumn,showpacs]{revtex4}

\usepackage{graphicx}
\usepackage{amsmath}
\usepackage{amssymb}

\begin{document}
\def\a{\alpha}
\def\b{\beta}
\def\e{\varepsilon}
\def\d{\delta}
\def\l{\lambda}
\def\m{\mu}
\def\t{\tau}
\def\n{\nu}
\def\o{\omega}
\def\r{\rho}
\def\S{\Sigma}
\def\G{\Gamma}
\def\D{\Delta}
\def\O{\Omega}

\def\ra{\rightarrow}
\def\ua{\uparrow}
\def\da{\downarrow}
\def\pd{\partial}
\def\bk{{\bf k}}
\def\bp{{\bf p}}
\def\bn{{\bf n}}

\def\be{\begin{equation}}\def\ee{\end{equation}}
\def\bea{\begin{eqnarray}}\def\eea{\end{eqnarray}}
\def\nn{\nonumber}
\def\lb{\label}
\def\pref#1{(\ref{#1})}


\title{Adiabatic theory of boundary friction and stick-slip processes}

\author{Yu.G.~Pogorelov}
\affiliation{CFP/Departamento de F\'{i}sica, Universidade do Porto, 4169-007 Porto,
Portugal}

\date{\today }

\begin{abstract}
An adiabatic approach is developed for the problem of boundary friction
between two atomically smooth and incommensurate solid surfaces, separated
by a monolayer of lubricant atoms. This method permits to consider very slow
macroscopic motion of the parts in contact, separately from fast thermic
motions of individual atoms. A characteristic ''stick-slip'' behavior of
the tangential force on the contact is obtained within a simple 1D model,
relevant for the tip and sample system in friction force microscopy (FFM).
This behavior reflects the specific mechanism of stress energy
accumulation, through formation of long-living metastable states (defects)
within the monoatomic lubricant layer, and their subsequent collapse with 
energy conversion into heat. This is similar to the dislocation mechanism of
irreversible deformation in bulk solids. The peculiar feature predicted by
the present theory is the twofold periodicity of ''stick-slip'' spikes with
relative displacement: the shorter period $a\delta $ (where $a$ is the tip
lattice periodicity and $\delta $ the relative tip-sample lattice mismatch)
relates to defect skips by one elementary cell, and the longer period 
$a(1-\delta )$ relates to defect annihilation or nucleation at the boundaries
of contact area.
\end{abstract}


\maketitle
\section{\protect\bigskip Introduction}
Atomic mechanisms underlying the phenomenon of friction attract now a still
increasing interest of many investigators. Particularly, this is motivated
by the need of better understanding the physical principles that govern the
image formation in friction force microscopy (FFM) \cite{mate}. Commonly, we
call friction the process of energy dissipation during the relative
displacement of two solid surfaces in contact \cite{bowden}. The friction
force $F(x)=-dW(x)/dx$, where $W(x)$ is the work performed to reach the
position $x$ (Fig. \ref{Fig.1}), is always directed contrarywise to the 
displacement. The empirical ''Amontons' law'' relating this force to the normal 
load $N$ on contact: $F=kN$, where the friction coefficient $k$ depends only on the
nature of the contacting materials, is known for about three centuries \cite{dowson}. 
The modern tribology much dealt with the linear relation between N
and the effective contact area A, through deformations of microscopic
rugosities \cite{tabor}, so that $F=S_{c}A$ with a certain load independent
constant $S_{c}$ (the Bowden-Tabor law).

\begin{figure}
\centering{
\includegraphics[width=6.cm, angle=0]{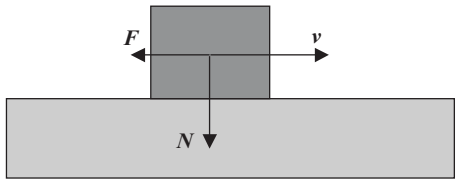}}
\caption{Friction force $F$ vs applied load $N$ at given velocity $v$ of relative 
displacement of two bodies.}
\lb{Fig.1}
\end{figure}

Nowadays, it is being progressively conceived that essential physics is
already contained in the fundamental process of friction between two
atomically smooth solid surfaces, not accompanied by any morphologic changes
(wearless friction). It is also well recognized that a rather necessary
premise for such a regime is the presence of some intermediate (lubricant)
layer of relatively inert atoms between the surfaces. Usually, this friction
regime is referred to as boundary friction \cite{bowden}. The most evident
source of dissipation at wearless friction is the generation, by microscopic
interactions at the contact, of various kinds of quasiparticles. Those
should bear away a part of mechanical energy from the external drive and
finally transform it into heat.

To the date, a plenty of experimental material on this sort of friction is
obtained. Particularly, a lateral force contrast was discovered in FMM scans
along an atomically smooth surface containing domains of different materials 
\cite{colchero,meyer}, thus showing the constant $S_{c}$ in the
Bowden-Tabor law to be material dependent. Furthermore, in FFM experiments
with the best resolution, the friction force vs time (or displacement) isn't
constant but reveals a sawtooth-like, or the so-called stick-slip, behavior 
\cite{mate}. This should indicate ocurrence of certain mechanical
instabilities during the displacement, either within the contact itself
or/and within the measurement apparatus (cantilevers, drives, etc.). The
theory related to cantilever instabilities \cite{tomanek,col} has its
main disadvantage in that it doesn't account for the irreversible
(microscopic) processes. The models for the contact instabilities, such as
the independent oscillator model \cite{tom,mcclel} or the
Frenkel-Kontorova model \cite{frenkel,sokoloff}, though being able to
explain the energy losses, were limited so far to only direct solid-solid
interactions (see also \cite{hirano}), not mediated by the lubricant. For the 
processes of boundary friction, the most elaborated approach is that considering 
the sequence of ``freezing-melting'' transitions within lubricant layers of few 
atoms thickness as a source of stick-slip discontinuities \cite{isr}. However, 
the FFM conditions can already reach to ultrathin boundary layers, monoatomic or 
even submonoatomic. In this situation, the interlayer viscosity and hence the 
melting transitions are most probably excluded, nevertheless the indication of 
still present discontinuous friction forces (atomic contrast in FMM) poses a 
challenge for atomic friction theory. The treatment in this paper just concerns 
the contact instabilities between two solids separated by a monoatomic lubricant 
layer and define microscopic mechanisms for irreversible energy losses with atomic 
periodicity at slow relative displacement. To fulfill this program, it is of special 
importance to choose an adequate calculation method.

The most straightforward method, very popular now for modelling various
processes on nanoscopic level, is that of molecular dynamics (MD), and it
was also applied to boundary friction \cite{landmann,nieminen,persson}. However, 
this method has one substantial limitation that was not yet addressed explicitly. 
The time step for integrating the MD equations is typically of order of femtoseconds, 
then, in order to meet reasonable computing time requirements (no more than 10$^{4}$-
10$^{6}$ steps), one has to consider the displacement velocities v not slower than 
$\sim 10^{3} $ cm/s to model the displacement by only few atomic periods. But in reality, 
even the faster process of atomic sliding of a free Lennard-Jones adsorbate layer on a 
single metal surface, subject to a uniform parallel force \cite{krim}, is 
characterized by much slower velocities $v\sim 0.1$ cm/s. And the velocities in 
STM, AFM, FFM, etc. techniques, scale from $10^{-4}$ down to $10^{-7}$ cm/s, thus 
opening a 10 orders of magnitude abyss beyond the MD capacities. Meanwhile, just 
such slow adiabatic motions, as will be also seen further, are characteristic for 
the observable friction phenomena.

The remedy for this problem can be found in focusing the treatment on
possible long living metastable states in the boundary layer, similar to
those, well-known in physics of solids, as dislocations, domain walls, etc.
Then the slow dynamics (SD) of such states can be considered separately from
fast thermal motions of MD. Adiabatically, the SD part just follows the
scenarios of contact instabilities of Refs. \cite{tom,mcclel,frenkel,sokoloff,hirano}, 
while the MD part is accounted for in average, through the temperature dependent factor,
triggering the transitions between different (meta)stable states. (This also
can be meant as if the adiabatic atomic potentials were defined only within
to $\sim \beta ^{-1}=T$.)

Each transition is preceded by a certain period when the energy of a
metastable state ({\it m}-state), and so the elastic force on the system,
is growing with the macroscopic displacement $x$, while the energy barrier $%
h $, separating the {\it m}-state from the nearest stable ({\it s}-)
state, is decreasing (Fig. \ref{Fig.2}). The adiabaticity condition would require that
the {\it m}-state lifetime, $\tau _{m}=\tau _{a}\exp (\beta h)$, does not
change too rapidly (see a more precise criterion at the end of Sec. 3). When 
$\tau _{m}$ decreases so much that it becomes comparable to the
characteristic time $\tau _{v}$ $=a/v$ of slow displacement by an atomic
period $a$ (though still at $\beta h\gg 1$), the barrier can be overcome by
a thermal fluctuation. Then, within a very short time $\sim \tau _{a}$, the
system passes from {\it m} to another state {\it s}*, highly excited 
by an energy $\varepsilon ^{\ast }\gg $ $\beta ^{-1}$ (see Fig. \ref{Fig.2}) over 
{\it s}. Next, within a certain relaxation time $\tau _{r}$ $\sim 10^{8}$ s, 
the energy difference $\varepsilon ^{\ast }$ is released through
emission of about $\beta \varepsilon ^{\ast }\gg 1$ quasiparticles.

\begin{figure}
\centering{
\includegraphics[width=6.cm, angle=0]{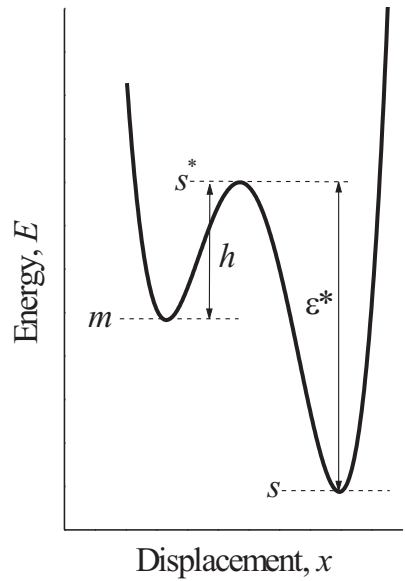}}
\caption{Energy level scheme for transition between a metastable {\it m}-state 
to a stable {\it s}-state, through an excited {\it s}*-state. The excitation 
energy is $\varepsilon ^{\ast }$ and the barrier height $h$.}
\lb{Fig.2}
\end{figure}

Since $\tau _{r}$ is much shorter of the characteristic time ($\sim \tau _{m}$) of 
the inverse s* $\rightarrow $\textit{m} transition, the latter possibility can be 
safely excluded \cite{note}. The considered \textit{m }$\rightarrow $\textit{s}*
$\rightarrow $ s process gives an elementary contribution to irreversible losses 
and can be compared to an individual spike in the stick-slip picture. The work by 
displacement $\Delta x$ is written, accordingly to the 1st law of thermodynamics, 
as 
\begin{equation}
\Delta W=\Delta E+\Delta Q  \label{eq1}
\end{equation}
Between the transitions, the heat transfer $\Delta Q=0$ and the force varies
continuously (the stick stage). Contrarywise, on a transition, the entire
energy change is released into heat: $\Delta E=-\Delta Q$ , hence $\Delta
W=0 $, and a force discontinuity is generated (a slip). In this first
attempt, the model is reduced to the simplified situation of 1D interface
between 2D ''solids'', however its extension to the realistic 2D (and
possibly curved) interface can be also done on the same conceptual basis.

\section{Formulation of the model}

Let us consider a system (Fig. \ref{Fig.3}) consisting of two 2D ''solid'' arrays:
conventionally called ''tip'' ({\it t}) and ''sample'' ({\it s}),
separated by an atomic chain of ''lubricants'' ({\it l}) at some distance 
$d$. We assume the same triangular lattice for both solids, but admit a
small mismatch $\delta $ between the lattice parameters. The {\it t}-side
of the contact consists of $L$ elementary cells ($L+1$ atoms), while the 
{\it s}-side is unlimited, and the {\it l}-chain in contact with 
{\it t}\ consists of $L$ atoms. Without loss of generality, we can put
the {\it t}-lattice parameter equal $a$ and the {\it s}-lattice
parameter $a(1-\delta )$, $\delta \ll 1$. 

\begin{figure}
\centering{
\includegraphics[width=8.cm, angle=0]{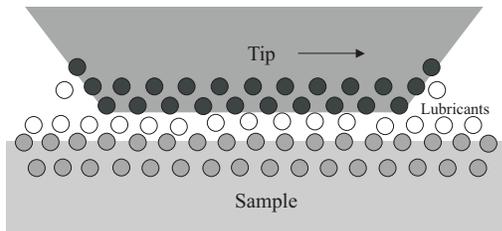}}
\caption{General view of the contact area. The dark circles represent tip 
atoms, the pale grey are sample atoms, and the empty circles are lubricant 
atoms.}
\lb{Fig.3}
\end{figure}

These lattice structures are supposed not to vary at bringing {\it t} and {\it s} 
into ({\it l}-mediated) contact and during their posterior mutual displacement by $x$.
Thus we assume the solid-solid links infinitely rigid compared to the
solid-lubricant and lubricant-lubricant links. Hence, the only variable part
in the full system energy is that related to the {\it l}-subsystem: 
\begin{equation}
E\left( x\right) =\sum_{n=1}^{L}\left[ U\left( \mathbf{r}_{n},x\right) +%
\frac{1}{2}\sum_{n^{\prime }\neq n}V\left( r_{n,n^{\prime }}\right) \right] ,
\label{eq2}
\end{equation}
where $U\left( \mathbf{r}_{n},x\right) $ is the ''mean-field'' potential on
the n-th lubricant with 2D radius-vector $\mathbf{r}_{n}$ for given {\it t}-
{\it s} displacement $x$, and $V\left( r_{n,n^{\prime }}\right) $ is
the {\it l}-{\it l} interaction ($\mathbf{r}_{n,n^{\prime }}=\mathbf{r}%
_{n}-\mathbf{r}_{n^{\prime }}$). Strictly speaking, the energy, Eq. (\ref
{eq2}), is also a function of all $\mathbf{r}_{n}$, but if we only admit
them to belong to stable (or metastable) {\it l}-configurations at a
fixed distance $d$, the displacement $x$ remains the single relevant
parameter. The specifics of the above formulation consists, firstly, in
taking into account two atomic periodicities (generally incommensurate) at
once and, secondly, in the bilateral restriction of lubricant layer between
two surfaces. Qualitative difference of this situation from the known
microscopic models of a lubricant layer over single solid surface with
single periodicity \cite{tom,mcclel,frenkel,sokoloff,hirano,persson} was already 
recognized in literature \cite{jacob}. For simplicity, we will construct all the 
interactions in Eq. (\ref{eq1}) using the standard Lennard-Jones potential: 
\begin{equation}
f_{\mathrm{LJ}}\left( y\right) =y^{-12}-y^{-6}.  \label{eq3}
\end{equation}
Particularly, we put 
\begin{eqnarray}
U\left( \mathbf{r}_{n},x\right) &=&\varepsilon _{0}\left[ \sum_{m} 
f_{\mathrm{LJ}} \left( \left| \mathbf{r}_{n}- \mathbf{r}_{m}\right| /r_{t}\right)
 \right. \notag \\
&& +\left. \sum_{m^{\prime }}f_{\mathrm{LJ}}\left( \left| \mathbf{r}_{n}-\mathbf{r}%
_{m^{\prime }}-\mathbf{x}\right| /r_{s}\right) \right] , \nonumber
\end{eqnarray}
where $\varepsilon _{0}$ is the adhesion energy, $\mathbf{r}_{m}$ and $%
\mathbf{r}_{m^{\prime }}$ are respectively the coordinates of {\it t}-
and {\it s}-atoms, $r_{t}$ and $r_{s}$ the corresponding {\it t}-%
\textit{l} and {\it s}-{\it l} equilibrium distances, and $\mathbf{x}$
the vector of displacement of the {\it t}-. Thus, both the {\it t}-%
{\it l} and {\it s}-{\it l} coupling energies equal $\varepsilon
_{0}$, serving as the energy scale. At least, the \textit{l}-\textit{l}
interaction is taken $V\left( r_{n,n^{\prime }}\right) =g\varepsilon _{0}f_{%
\mathrm{LJ}}\left( r_{n,n^{\prime }}/r_{t}\right) $, where $g$ is the
coupling constant and $r_{l}$ the \textit{l}-\textit{l} equilibrium distance.

\section{Mean-field potential and metastable states}

One can reasonably consider the \textit{l}-\textit{l} interaction the
weakest one in the above system and, at the first step, put $g=0$ (this is a
good approximation unless the lubricants approach each other too closely
compared with $r_{l}$). Then the equilibrium states will correspond to
various distributions of $L$ lubricants over the minima of the mean-field
potential produced by the \textit{t}- and \textit{s}-lattices. In what
follows we suppose that the infinite sample moves with velocity $v$ in $x$%
-direction with respect to the fixed tip, and the tip-sample distance $d$ is
constant. We define the $n$-th elementary cell of the boundary layer as the
rectangle limited by the $n$-th and $(n+1)$-th tip atoms in $x$-direction,
and $d$ wide in $y$-direction, and the entire contact area corresponds to $%
1<n<L$ (inset to Fig. \ref{Fig.4}).

\begin{figure}
\centering{
\includegraphics[width=6.cm, angle=0]{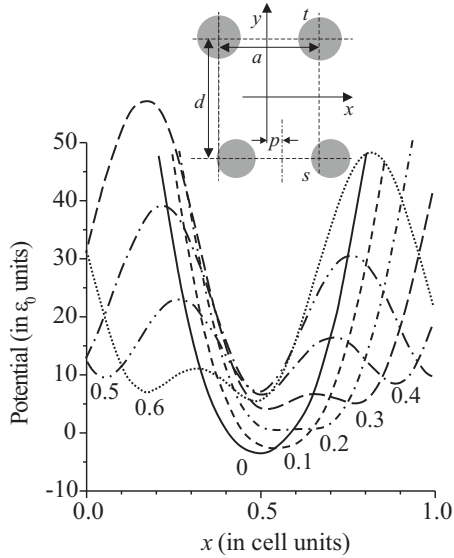}}
\caption{Mean-field potential profiles for a lubricant atom in the unit cell of 
a contact, at different values of the phase variable $p$ (marks at each curve). The 
mismatch parameter $\delta = 0.05$, the interfacial separation $d = 1.12 a$. 
Inset: the phase variable $p$ is defined as the distance along $x$-coordinate 
between the midpoints of nearest neoghbor \textit{t}- and \textit{l}-atoms within 
the cell.}
\lb{Fig.4}
\end{figure}

Let the initial tip-sample relative position $x=0$ at the instant $t=0$ be
such that the \textit{s}-atoms are located symmetrically in the zeroth cell.
In what follows, we consider the potential energy $U_{n}$ of a lubricant in $%
n$-th cell as a function of $x_{n}$ only (as shown in Fig. \ref{Fig.4}), supposing 
$y_{n}$ be always adjusted to the relative minimum of this energy at given 
$x_{n}$ (i.e., supposing the lubricants only to move along energy ''valleys''
in 2D cells). Because of the small mismatch $\delta $, the static potential
relief $U_{n}\left( x_{n}\right) $ ($0\leq x_{n}\leq a$) slowly varies (in
general, incommensurably) from cell to cell along the interface so that the
cells aren't equivalent. If we neglect the boundary effects (as is the case
in what follows), the potential relief within $n$-th cell is fully determined by 
the ''phase parameter'' $p_{n}=x-n\delta (\mathrm{mod}(1-\delta ))$.

As seen from Fig. \ref{Fig.4}, besides the general raising or lowering of minima 
with $p$ (elastic strains), there also exist some critical values: $p_{c1}(\approx
0.22$ in that particular case) and $p_{c2}=1-\delta -p_{c1}$, when the
initial single minimum $U^{(0)}$ splits into two. In the splitted potential,
the lower minimum $U^{(1)}$ is separated from the upper one $U^{(2)}$ by a
maximum $U^{(3)}$ (a saddle point in the 2D cell). The slow relative \textit{t}-\textit{s} 
displacement will result in that the potential relief as if ''moves'' adiabatically along 
$x$ with the ''phase velocity'' $v/\delta $.

If all the lubricants always rest in the lower minima $U^{(1)}$, the
distances between them never become too short in this process and we can
still neglect the \textit{l}-\textit{l} interactions. Then no irreversible
losses and, hence, no friction occur in the system (the force on external
drive $F=-dE(x)/dx$, though non-zero, results fully elastic). However, such
an \textit{l}-configuration, even at moderate $L$ (note, however, that a
contact area in FMM typically includes $\symbol{126}10^{3}$ or more
elementary cells), cannot persist for any macroscopic time by the
thermodynamic reasons.

Indeed, with growing displacement, the cell potential splits and then the
barriers $h_{1}=U^{(3)}-U^{(1)}$ and $h_{2}=U^{(3)}-U^{(2)}$ begin to grow
and eventually get practically impenetrable. Then there is a finite
probability $w>0$ that at this stage a lubricant atom leaves ''captured'' in
the upper, metastable minimum $U^{(2)}$. For instance, in the model of
linearly growing barriers with time (see Appendix A), this probability is
simply given by the Fermi distribution function for the two-level system: 
\begin{equation}
w\left( \Delta E\right) =\left( \mathrm{e}^{\beta \Delta E}+1\right) ^{-1}
\label{eq4}
\end{equation}
being $\Delta E$ the energy difference between the two minima. If all $L$
lubricants are initially in $U^{(1)}$ states, the time necessary for at
least one of them pass to $U^{(2)}$ is about $1/(Lwv)$, which corresponds to
the system displacement by only $1/(Lw)$ atomic periods while $Lw$ can
easily exceed unity.

\begin{figure}
\centering{
\includegraphics[width=11.cm, angle=0]{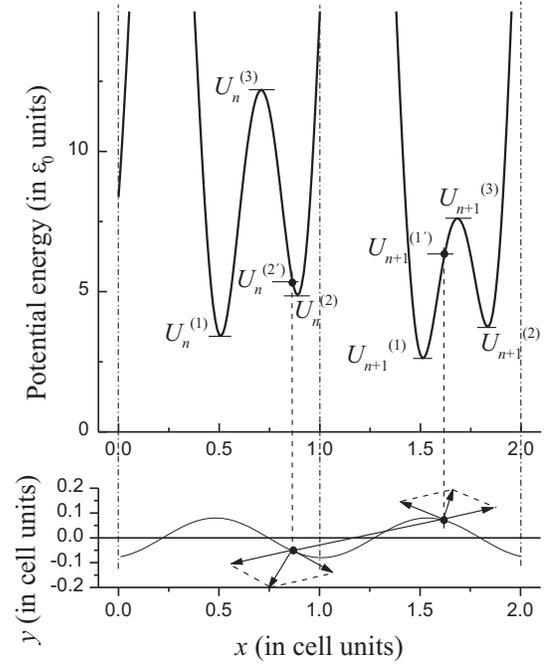}}
\caption{Relative positions of two interacting lubricant atoms, $n$ and $n + 1$,
against the mean-field potential (upper part) and the valley of steepest
descent in $xy$-plane (lower part). The mean-field potential is that of Fig.
\ref{Fig.4} and the interaction constant $g = 1/120$. The phase $p_n$ is close to 
the transition value $p_t$.}
\lb{Fig.5}
\end{figure}

Now let the $n$-th lubricant leave ''captured'' at $U_{n}^{(2)}$, the upper
minimum of $n$-th cell, while its neighbor in $(n+1)$-th cell stays at 
$U_{n+1}^{(1)}$, and follow the system evolution with displacement (Fig. \ref{Fig.5}).
As the distance $r_{n,n+1}$ gets shorter, this particular \textit{l}-\textit{%
l} interaction cannot be more neglected. Such a pair of strongly interacting
lubricants just realizes a metastable state for the considered system. The
Lennard-Jones repulsion will slightly shift the $n,(n+1)$ pair from the
mean-field minima $U_{n}^{(2)}$ and $U_{n+1}^{(1)}$ towards the saddle
points $U_{n}^{(3)}$ and $U_{n+1}^{(3)}$. If the interaction constant $g$
isn't too small (for the particular case considered in Sec. 4, $g$ = 1/120
is already sufficient), such repulsion makes the energy barrier between this 
\textit{m}-state and the \textit{s}-state to turn from growing to lowering.
The displaced equilibrium positions $U_{n}^{(2^{\prime })}$ and $U_{n+1}^{(1^{\prime })}$ 
are defined by the compensation between the \textit{l}-\textit{l} interaction and 
the crystalline field forces along the valleys: 
\begin{eqnarray}
\mathbf{\Delta }_{n}U_{n}\cdot \left( \mathbf{\Delta }_{n}U_{n}+\mathbf{%
\Delta }_{n}U_{n,n+1}\right) &=&0,  \label{eq5} \\
\mathbf{\Delta }_{n+1}U_{n+1}\cdot \left( \mathbf{\Delta }_{n+1}U_{n+1}+%
\mathbf{\Delta }_{n+1}U_{n,n+1}\right) &=&0,  \notag
\end{eqnarray}
($\mathbf{\Delta }_{n}$ stands for the 2D gradient along $\mathbf{r}_{n}$).
Notably, the numeric results (see Fig. \ref{Fig.5} and Sec. 4) show that the energy
difference $U_{n+1}^{(1^{\prime })}-U_{n+1}^{(1)}$ for the $(n+1)$-th
lubricant increases much faster than $U_{n}^{(2^{\prime })}-U_{n}^{(2)}$ for 
$n$-th one. Hence, as the repulsion grows with $x$, the barrier 
$h_{n+1}^{(1)}=U_{n+1}^{(3)}-U_{n+1}^{(1^{\prime })}$ decreases, while 
$h_{n}^{(2)}=U_{n}^{(3)}-U_{n}^{(2^{\prime })}$ remains increasing, and the 
$(n+1)$-th atom will be finally pushed out by a thermal fluctuation from the 
$U_{n+1}^{(1^{\prime })}$ position to $U_{n+1}^{(2)}$. Then the distance 
$r_{n,n+1}$ will suddenly increase and therefore the $n$-th atom will relax
from $U_{n}^{(2^{\prime })}$ to $U_{n}^{(2)}$. The overall energy gain 
\begin{equation*}
\Delta \varepsilon =U_{n+1}^{\left( 1^{\prime }\right) }+U_{n}^{\left(
2^{\prime }\right) }-U_{n}^{\left( 2\right) }-U_{n+1}^{\left( 2\right) }
\end{equation*}
(not to be confused with the energy difference $\Delta E$ in Eq. (\ref{eq4})) is 
released through emission of quasiparticles. Supposing isotermic conditions, this 
energy is eventually transferred from the external drive to the thermostate.

Further on, the $n$-th lubricant stays always at $U_{n}^{(2)}$ (until it
returns to the unsplitted position $U_{n}^{(0)}$ at $p_{n}\rightarrow p_{c2}$) while 
the metastable pair is now formed by the $(n+1)$-th and $(n+2)$-th
lubricants (that is, skipped by one cell) and so on. For the hierarchy of
times specific for our system, the transition time $t_{tr}$ (related to the
corresponding phase value $p_{tr}$ through $p_{tr}=vt_{tr}$) has a very
small statistical dispersion around its mean value $\left\langle
t_{tr}\right\rangle $ (see Appendix B). Also the mismatch parameter $\delta $
can be chosen so that either the ''capture'' and ''skipping'' processes
affect only single metastable pair at a moment, not its nearest neighbors,
thus avoiding possible complications due to interacting fluctuations.

Once randomly initiated, each skipping process described above provides a
rather regular and stable generator of energy losses within a ''phase
domain'' of length $l_{d}\approx a(1-\delta )/\delta $ in the \textit{l}-chain. 
Along such a domain, the phase $p$ changes from 0 to $1-\delta $,
and the lubricants are in the $U^{(2)}$ position at all the sites with 
$p_{n}<\left\langle p_{tr}\right\rangle =v\left\langle t_{tr}\right\rangle $,
and in $U^{(1)}$ position at $p_{n}>\left\langle p_{tr}\right\rangle $ (for
more details see in \ref{Fig.6}). The spikes in the friction force from
subsequent skips have approximate periodicity $\approx \delta a/v$ in time,
or $\approx \delta a$ in displacement. Since this is much smaller of atomic
periodicity, it may be difficult to resolve such fine structure in actual
experiments.

\begin{figure}
\centering{
\includegraphics[width=8.cm, angle=0]{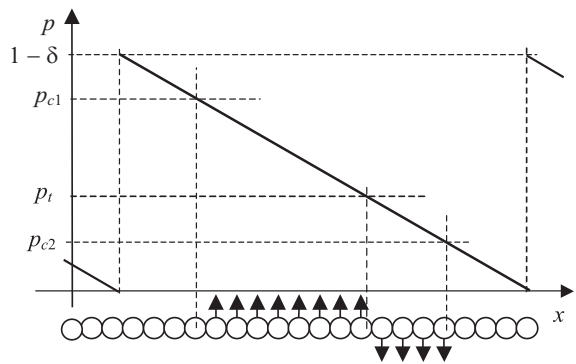}}
\caption{Domain structure of the lubricant chain formed at displacement of
the contacting solids. The "spin" representation is chosen, where the spin-up
corresponds to an atom in upper well, spin-down to that in lower well, and
spinless is an atom in unsplitted well.}
\lb{Fig.6}
\end{figure}

But the atomic periodicity itself can result from another source. The
described generators move along the chain with phase velocity $v_{p}=v/\delta $ and, 
at sufficient distances from the beginning of chain ($n=1$), they exist within each 
domain, $l_{d}$ apart. The friction force, averaged over the $\delta $-periods, is 
$\left\langle F\right\rangle _{\delta }\approx n_{g}\Delta \varepsilon /(\delta a)$, 
where $n_{g}$ is the total number of generators in the chain. When a generator reaches 
the chain end ($n=L$), it disappears (like a dislocation when it reaches the sample
surface) and $n_{g}$ decreases by unity. Hence a jump to down occurs in the average 
force $\left\langle F\right\rangle _{\delta }$. Otherwise, when a new generator
appears near the chain beginning, $\left\langle F\right\rangle _{\delta }$
increases. These events have the periodicity $\approx l_{g}/v_{p}=a(1-\delta
)/v$ in time, that is just the sample cell periodicity $a(1-\delta )$ in
displacement.

The averaged friction force $\left\langle F\right\rangle $ over many atomic
periods results simply $\left\langle F\right\rangle \approx L\Delta
\varepsilon $, that is proportional to the ''contact area'' $L$ with the
Bowden-Tabor coefficient $\Delta \varepsilon $, which depends both on the
material parameters and the interfacial distance $d$ (related to the
pressure).

To conclude this section, we obtain a quantitative criterion for
adiabaticity in our system. The limitation on the displacement velocity $v$
can be extracted from the evident condition that the lifetime $\tau _{m}$ of
the metastable state (especially when it becomes comparable to $\tau _{v}$)
doesn't changes appreciably within the microscopic time $\sim \Omega ^{-1}$
(atomic oscillation period): $d\tau _{m}/dt\ll \Omega \tau _{m}$, or $\beta
dh/dt\ll \Omega $. Since $dh/dt=vdh/dx$ and the barrier height changes from
its maximum value $h_{max}$ to zero within a fraction $q\ll 1$ of atomic
period, the adiabaticity criterion is given by: 
\begin{equation}
v\ll v_{cr}=\frac{a\Omega }{\beta h_{max}/q}  \label{eq7}
\end{equation}
The numerator in the last equation is of the order of sound velocity while
the denominator $\beta h_{max}/q$ can be as high as $\sim 10^{2}$.
Thus, the adiabaticity must be definitely violated (with all possible
complications, as appearance of shock waves, thermal instabilities, etc.) at 
$v$ above some m/s, as is the case for the most of MD simulations. However,
it is well assured at actual FFM velocities, mentioned above.

\section{Numerical results}

We choose the equilibrium distances for Lennard-Jones interactions $r_{t}=0.8a$, 
$r_{s}=0.78a$, $r_{l}=0.9a$, and the mismatch parameter $\delta =0.05$. The energy 
parameter $\varepsilon _{0}$ is chosen 10 meV and the cell parameter $a = 0.3$ nm. 
At least, the interfacial distance is chosen $d=1.12a$. This value is slightly 
below the equilibrium separation $d_{e}=1.24a$ for a cell at the displacement 
phase $p=0$ and corresponds to the normal force on tip $N=17.2$ nN. The calculated 
profiles of mean-field potential for different values of the phase parameter $p$ are 
shown in Fig. \ref{Fig.4}. Next, the equilibrium equations, Eq. (\ref{eq5}), are resolved 
to give the values of barriers $h^{(1)}$ and $h^{(2)}$ as functions of $p$. The total
length of the system was taken $L=20$ and its initial state corresponds to
one lubricant at $U^{(1)}$ (or $U^{(0)}$) state in each cell. The
displacement velocity $v=30$ nm/s   corresponds to the characteristic time
of displacement by cell period $t_{v}=10^{-2}$ s, and the elementary time
step is $\Delta t=10^{-5}$ s, that is 10 orders of magnitude slower of the
respective MD times. The evolution of total energy $E(x)$ is modelled,
using the mean-field and \textit{l}-\textit{l} interaction energies (at
displaced positions for metastable pairs) at phase values $p_{n}$ which vary
with $x=vt$, and the account for capturing and skipping events with
probabilities is described in Appendices A,B. The temperature is set $T=1$ meV 
$\approx $ 11 K.

\begin{figure}
\centering{
\includegraphics[width=8.cm, angle=0]{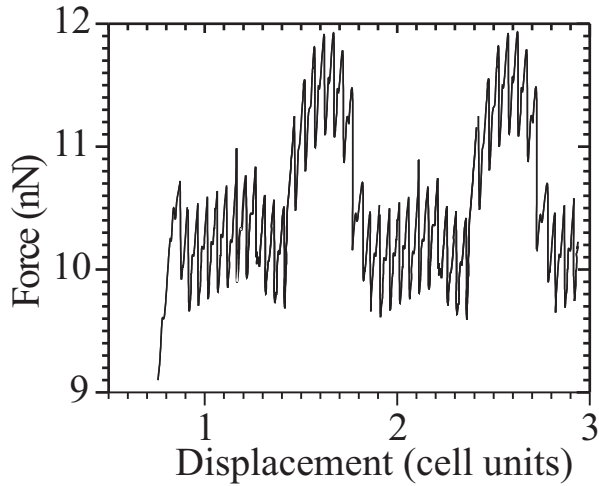}}
\caption{Microscopic tangential force on the contact vs sample displacement.
Each fine structure spike corresponds to a skip of the boundary between
"spin-up" and "spin-down" areas of Fig. \ref{Fig.6} by one cell period, while the
"envelope sawtooth" with the periodicity of sample lattice (0.95a) corresponds to 
the disappearings of domain boundaries at reaching the sample end.}
\lb{Fig.7}
\end{figure}
The resulting plot of the friction force $F(x)$ is presented in Fig. \ref{Fig.7}.
After some initial period, when the force is almost elastic, the stick-slip
behavior with two types of periodicity establishes and the mean level of
friction force corresponds to a rather high ''friction coefficient'' $k\approx 0.6$. 
A more detailed study of the present model, for different values of pressure, 
temperature and velocity parameters, will be given elsewhere. It is also of interest, 
that using the above parameter values and $\Omega =10^{12}$ s$^{-1}$ in the adiabaticity 
criterion, Eq. (\ref{eq7}), we get the critical velocity $v_{cr} \sim 0.3$ m/s, in good 
accordance with our previous reasoning.

\section{Concluding remarks}

The above analysis gives only a simple illustration for possible irreversible processes 
within the adiabatic picture of boundary friction. Other types of the system behavior 
can be obtained at varying the material and external parameters. In particular, with 
growing separation $d$ between surfaces (lower or even negative pressure) the potential 
profiles get lower and the stick-slip behavior is expected to disappear. Contrarywise, at
closer separation, the stage of inverse barrier evolution ceases to appear
for lubricant pairs, and irreversible transitions (much sharper) can occur
already in close clusters of three or more lubricants, however this would
greatly complicate the treatment. A special regime can be obtained for the
case of two fully commensurate solids ($\delta =0$). In this case, there is
no preference between the two splitted mean-field states in a cell, and the
''phase domain'' becomes infinitely long. All the skips would occur
simultaneously over the whole contact area, at the moment when the
transition phase value ptr is reached at all the cells. This will generate
very high and short spikes of friction force with single periodicity of the
lattice. Note in conclusion that passing from the above considered 1D to the
realistic 2D situation will complify the model, not only by the extension of
arrays for computation, but also due to introduction of several types of
different metastable states.

\section{Appendix A}

To model the initial evolution stage of potential in our system, let us
consider a particle in the double-well system (Fig. \ref{Fig.8}) where the 
potential barriers for each well $h_{1,2}$ vary adiabatically with time at 
$t\geq 0$.
 
\begin{figure}
\centering{
\includegraphics[width=6.cm, angle=0]{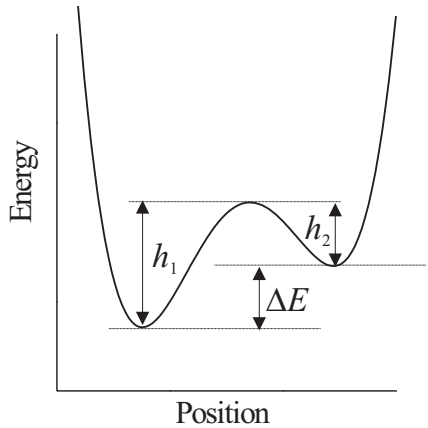}}
\caption{Double-well potential for an isolated lubricant atom.}
\lb{Fig.8}
\end{figure}

Then the probability $w(t)$ to find the particle in the upper well 
at the moment $t$ obey the linear differential equation with the coefficients
slowly depending on time (through $h_{1,2}\left( t\right) $):

\begin{equation}
\Omega ^{-1}\frac{dw}{dt}=-\left( \mathrm{e}^{-\beta h_{1}}+\mathrm{e}%
^{-\beta h_{2}}\right) w+\mathrm{e}^{-\beta h_{1}}.  \tag{A1}  \label{a1}
\end{equation}
The solution to Eq. (\ref{a1}), satisfying the initial condition $w(0)=0$
(that is, at $t=0$ the particle is with certainty in the lower well), is
given by the evident expressions: 
\begin{equation}
w\left( t\right) =\Omega \mathrm{e}^{-\varphi \left( t\right) }\int_{0}^{t}%
\mathrm{e}^{-\varphi \left( t^{\prime }\right) -\beta h_{1}\left( t^{\prime
}\right) }dt^{\prime }
\end{equation}
where
\begin{equation}
\varphi \left( t\right) =\Omega \int_{0}^{t}\left[ \mathrm{e}^{-\beta
h_{1}\left( t^{\prime }\right) }+\mathrm{e}^{-\beta h_{1}\left( t^{\prime
}\right) } \right] dt^{\prime }.  \tag{A2}  \label{a2}
\end{equation}
\newline
Now let us adopt, for simplicity, the model linear law for the growth of the
barriers: 
\begin{equation}
h_{1}\left( t\right) =\Delta E+h_{2}\left( t\right) ,\ \ \ \ \ \ h_{2}\left(
t\right) =ut.  \tag{A3}  \label{a3}
\end{equation}
As seen from Fig. \ref{Fig.4}, Eq. (A.3) is a rather plausible approximation. Then
the explicit integration in Eq. (A.2) gives the saught probability: 
\begin{equation}
w\left( t\right) =\frac{1-\exp \left[ -\Omega \left( 1-\mathrm{e}^{-\beta
ut}\right) /\beta u\right] }{\mathrm{e}^{-\beta \Delta E}+1}  \tag{A4}
\label{a4}
\end{equation}
Since the rate of barrier growth is $u=vh_{max}/(aq)$, we conclude from
comparison with Eq. (\ref{eq7}) that the dimensionless parameter $\Omega/\beta u$ is just 
the adiabaticity ratio $v_{cr}/v$ and, for typical FFM conditions, its value is
enormous: $\sim 10^{8}$. Thus the exponential in the numerator of Eq. (A.4) rapidly 
vanishes and we arrive at the result of Eq. (4).

\section{Appendix B}

Let us turn to the stage of reversed evolution (see the paragraph before Eq.
(5)), and let the potential barrier (the lower one in Fig. \ref{Fig.8}) adiabatically
decrease in time as $h(t)=-ut$, $-\infty <t<0$. If at the moment $t$ the particle is 
in the well, the probability that it escapes within an infinitesimal time interval 
$dt$ is $dw(t)=\Omega \mathrm{e}^{\beta ut}dt$. Then the probability distribution 
$\rho (t)$ for the transition to occur at the moment $t$ is the product of $dw(t)/dt$ 
times the probability to survive within the well during the period from $-\infty $ to $t$: 
\begin{eqnarray}
\rho(t)&=&\frac{dw}{dt} \exp \left[ -\int_{-\infty }^{t}dw\left( t^{\prime }\right) \right] 
 \notag \\
&&= \exp %
\left[ -\frac{\Omega }{\beta u}\mathrm{e}^{\beta ut} -\beta u t \right].  \nonumber
\end{eqnarray}
 
\begin{figure}
\centering{
\includegraphics[width=8.cm, angle=0]{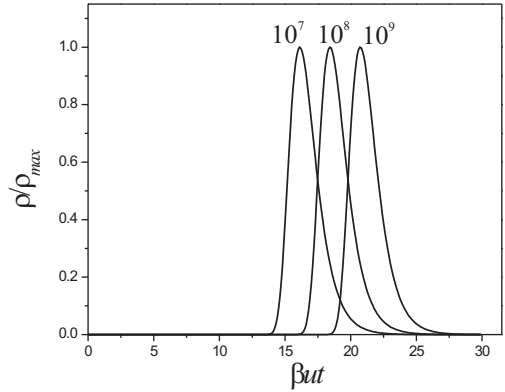}}
\caption{Probability distribution $\rho$, normalized to its maximum value $\rho_{max}$, 
as a function of dimensionless variable $\beta u t$ (see text) at different values of the
parameter $v_c/v$ (indicated at each maximum). The peak is quite narrow, its position 
varies very slightly over two orders of magnitude variation of $v_c/v$, while its form is 
just invariable.}
\lb{Fig.9}
\end{figure}

The function $\rho (t)$, shown in Fig. \ref{Fig.9} for different values of $u$,
sharply peaks at $t^{\ast }=\ln (\beta u/\Omega )/\beta u$. Since the
parameter $\Omega /\beta u=v_{cr}/v$ is huge \cite{note2}, then: i) the
distribution $\rho (t)$ is very narrow and ii) its form and the critical
barrier value $h^{\ast }=h(t^{\ast })=\beta ^{-1}\ln (v_{cr}/v)$ depend on
the displacement velocity $v$ only very weakly.


\begin{thebibliography}{99}
\bibitem{note2}  In fact, at this reverse evolution, when the potential
barrier is formed with essential participation of the Lennard-Jones
repulsion, the precise values of $dh/dt$ and hence of $v_{cr}$ somewhat differ from
those in Eq. (\ref{eq7}), for purely mean-field potential. However, the numerical
analysis shows that here the adiabaticity conditions hold even better.
\end{thebibliography}
\end{document}